\UseRawInputEncoding
\documentclass[12pt]{article}
\usepackage[vmargin=1.8cm, hmargin=3.2cm]{geometry}

\RequirePackage[OT1]{fontenc}
\usepackage{amsthm,amsmath,amssymb,enumitem,authblk,bm,graphicx}

\usepackage[english]{babel}  

\usepackage{tikz}
\usetikzlibrary{arrows.meta}

\usepackage{setspace}
\usepackage[round]{natbib}

\usepackage{tikz}
\tikzset{>={Latex[width=2mm,length=2mm]}}

\newcommand{\E}{\mathrm{E}}
\newcommand{\R}{\mathbb{R}}

\newcommand{\UG}{\mathcal{G}=(\mathcal{V},\mathcal{E})}
\newcommand{\re}{\mathbb{R}}

\newcommand{\ie}{{\it i.e.}}

\newcommand{\logit}{\mbox{logit}}
\newcommand{\logitInv}{\mbox{logit}^{-1}}

\newcommand{\Fnine} {\fontsize{9}{11}\selectfont  }
\newcommand{\Ften} {\fontsize{10}{11}\selectfont  }

\renewenvironment{abstract}
{\small
  \begin{center}
  \bfseries \abstractname\vspace{-.5em}\vspace{0pt}
  \end{center}
  \list{}{
    \setlength{\leftmargin}{.0cm}%
    \setlength{\rightmargin}{\leftmargin}%
  }%
  \item\relax}
{\endlist}

\usepackage[section]{placeins}

\usepackage{booktabs, siunitx}

\begin{document}

\pagestyle{myheadings}
\markboth{J.S. Pelck, H. Holthusen, M. Edelenbos, A. Luca and R. Labouriau}
{J.S. Pelck, H. Holthusen, M. Edelenbos, A. Luca and R. Labouriau}
\thispagestyle{empty}

\title{\Large
Multivariate Methods for Detection of Rubbery Rot  
 in Storage Apples by Monitoring Volatile Organic Compounds: \\ \large
An Example of Multivariate Generalised Mixed Models
}

\author[1] {Jeanett S. Pelck}
\author[2,3] {Hinrich H.F. Holthusen}
\author[3] {Merete Edelenbos}
\author[3] {Alexandru Luca}

\author[1,] {Rodrigo Labouriau 
  \thanks{Corresponding author: Rodrigo Labouriau, rodrigo.labouriau@math.au.dk}}
\affil[1]{Department of Mathematics, Aarhus University, Denmark}
\affil[2]{Esteburg Fruit Research and Advisory Centre, Germany}
\affil[3]{Department of Food Science, Aarhus University, Denmark}
\date{June 2021}

\clearpage\maketitle
\thispagestyle{empty}

\begin{abstract}\Ften
This article is a case study illustrating the use of a multivariate statistical method 
for screening potential chemical markers for early detection of post-harvest 
disease in storage fruit. We simultaneously measure a range of volatile organic 
compounds (VOCs) and two measures of severity of disease infection in apples under storage: the number 
of apples presenting visible symptoms and the lesion area. We use multivariate 
generalised linear mixed models (MGLMM) for studying association patterns of 
those simultaneously observed responses via the covariance structure of random 
components. Remarkably, those MGLMMs can be used to represent patterns of 
association between quantities of different statistical nature. In the particular 
example considered in this paper, there are positive responses (concentrations of 
VOC, Gamma distribution based models), positive responses possibly containing 
observations with zero values (lesion area,
Compound Poisson distribution based models) and binomially distributed 
responses (proportion of apples presenting infection symptoms).
We represent patterns of association inferred with the MGLMMs using graphical 
models (a network represented by a graph), which allow us to eliminate spurious 
associations due to a cascade of indirect correlations between the responses.
\end{abstract}


\normalsize

\newgeometry{top=3.5cm, bottom = 2cm, right = 3cm, left = 3cm}




\section{Introduction}

Rubbery rot is a post-harvest disease in apples caused by the fungus {\it Phacidiopycnis washingtonensis}, leading to significant storage losses in commercial production \citep{Ali2018}. Therefore, it is interesting to find predictors of rubbery rots onset at early stages of the infection development under fruit storage. To this purpose, a comprehensive study involving the emission of a range of specially chosen volatile organic compounds (VOCs) under apple storage conditions was performed by \cite{Holthusen2021a}. In this study, experimentally induced rubbery rots infections were monitored and contrasted with the concentration of 14 VOCs along the development of the disease, aiming to find chemical predictors for rubbery rot. This article exposes details of some non-standard statistical tools used in \cite{Holthusen2021a}.  

The experiment we will model can be shortly described in the following way.  Ten glass jars (below referred as glasses), containing nine inoculated apples were observed at three fixed observation times ($6$, $12$ and $18$ weeks post-inoculation). The following quantities were determined at each observation time: the number of apples presenting visible symptoms, the area of lesions caused by the fungal infection, and the air concentration of $14$ VOCs. The details of the experiment setup and the choice of the VOCs are exposed in \cite{Holthusen2021a}, see also \cite{Holthusen2021b} and \cite{Holthusen2021c}.

The proper statistical modelling of the complex of experiments referred to above presents several challenges. Indeed, the simultaneously observed responses are of different statistical nature. For example, while the number of apples showing visible symptoms (used to monitor the disease development) is naturally binomially distributed, the VOC concentrations follow continuous positive non-Gaussian distributions with high skewness. Furthermore, the area of lesions (characterising disease severity) presents many zero values (absence of infection) but otherwise follows a continuously skewed distributed and therefore is not adequately described by purely continuous distributions. Thus, the first challenge we encountered was to develop methods for establishing associations between these responses of different nature. We propose to solve this problem by using suitably constructed multivariate generalised linear mixed models (MGLMMs) simultaneously describing the responses referred to above. In this way, we will model the concentrations of the VOCs using Gamma distributions, representing positive valued responses with different degrees of skewness. Moreover, the lesion area will be modelled using Gamma compound Poisson distributions with positive mass at zero and otherwise continuous with variable degrees of skewness. All these families of distributions are particular cases of dispersion models, which are the families of distributions that form the basis of generalised linear mixed models (GLMMs). 

The MGLMM we propose to use is composed of marginal GLMMs describing each of the responses studied. Each of those GLMMs will contain a random component representing the basic experimental unit (the glass), which we use to model the covariance structure of the different responses. We use the tools of graphical models to describe this covariance structure in a suitable compact form, which will allow us to draw valid general conclusions on the association between the responses, even though they are of different statistical nature. For instance, we will eliminate spurious correlations between the responses, i.e., correlations between two responses that can be explained by a cascade of correlations between those responses and the other responses in play. In this way, we will identify a minimal group of VOCs sufficient to predict the rubbery rots onset, avoiding redundancy and the pitfalls of multicollinearity.

The MGLMMs we construct allow for incorporating corrections for determining factors know to have a strong influence in the responses (e.g., the observation week) and temporal correlation due to repeated observations at the same experimental unit. 

This paper is structure as follow. Section \ref{Section.2} introduces the marginal GLMMs for each of the responses considered. Those models are used to construct a MGLMM
 in Section \ref{Section.3}. The details of the construction are given in Section \ref{SubSection.3.1}, and the graphical model representing the covariance structure of the random components is discussed in section \ref{SubSection.3.2}. 
 Section \ref{Section.4}  briefly discusses the results obtained. 
 

\section{Models for Several Responses with Different Nature}
\label{Section.2}

We introduce below a range of GLMMs describing each of the observed responses. Those models contain a random component representing the glass (which is viewed as the basic experimental unit) and a fixed effect representing the observation time (week). We used a GLMM defined with the binomial distribution and logistic link function for describing the number of apples presenting visible symptoms. The concentrations of VOCs were modelled using GLMMs defined with a Gamma distribution and the logarithmic link function. Finally, we used a GLMM defined with the family of Gamma compound Poisson distributions and a logarithmic link to describe the lesion area. We give the full details of those models below using a notation suitable for defining the multivariate model for simultaneously describing the $16$ responses in play.

\subsection{Concentrations of Volatiles Organic Compounds - Positive Responses}\label{SubSection.2.1}

We describe below the GLMM used for modelling the concentration of each of the $14$ VOCs. We label those VOCs by the index $j$ ($j=1,\ldots,14$), which is kept fixed along this section (referring to a choice of one of the VOCs). 
Denote by $X_{tg}^{[j]}$ the random variable representing the concentration of the 
value of the $j^{th}$ VOC 
measured at the $t^{th}$ week ($t=6,12,18$) in the $g^{th}$ glass ($g=1,\ldots,10$).
According the GLMM we are defining, there exist $10$ unobservable 
random variables, denoted by $U^{[j]}_1,\ldots,U^{[j]}_{10}$, such that for $t=6,12,18$ and $g=1,\ldots,10$ the response  $X_{tg}^{[j]}$ is conditional Gamma distributed given $U_{g}^{[j]}$ 
with conditional expectation given by
\begin{align*}
	\log(\E[X_{tg}^{[j]}\vert U_{g}^{[j]}=u]) &= \theta_{t}^{[j]}+u \quad \text{ for all } u\in \R.
\end{align*}
Moreover, according to the GLMM the responses, 
$X_{gt}^{[j]}$ for $t=6,12,18$ and $g=1,\ldots,10$,  are conditionally independent given $U^{[j]}_1,\ldots,U^{[j]}_{10}$.
The specification of the GLMM is completed by stating that  $U^{[j]}_1,\ldots,U^{[j]}_{10}$ are independent and identically normally distributed with expectation $0$.
Here $ \theta_{6}^{[j]},  \theta_{12}^{[j]}$ and $ \theta_{18}^{[j]}$ are fixed effects describing the variation of the concentration of the $j^{th}$ VOC at different observation times.


\subsection{Number of apples Presenting Symptoms - Binomial Counts}
\label{SubSection.2.2}
Let $Y_{tg}$ be a random variable representing the number of apples presenting 
symptoms in the $g^{th}$ glass ($g=1,\ldots,10$), at the at the $t^{th}$ week  
($t=6,12,18$) out of the nine apples contained in each glass. We assume that there exist $10$ unobservable   random variables, denoted by
$V_1,\ldots,V_{10}$ such that, for $t=6,12,18$ and $g=1,\ldots,10$, $Y_{tg}$ is conditionally binomially distributed given $V_{g}$ with size $9$ and conditional expectation given by
\begin{align*}
	\logit(\E[Y_{tg}\vert V_{g}=v]) &= \alpha_{t}+v\quad \text{ for all } v\in \R \, .
\end{align*}
According to the model the responses, $Y_{gt}$ for $t=6,12,18$ and $g=1,\ldots,10$, are conditionally independent given $V_1,\ldots,V_{10}$. Moreover, the random components $V_1,\ldots,V_{10}$ are assumed to be independent and identically normally distributed with expectation zero.

\subsection{Lesion Area of Infection - Positive Responses with Zero Values}
\label{SubSection.2.3}
Denote by $Z_{tg}$ the random variable describing the observed lesion area in the $g^{th}$ glass ($g=1,\ldots,10$), at the $t^{th}$ week  
($t=6,12,18$). We assume that there exist $10$ independent and normally distributed random variables with expectation zero, denoted by $W_1,\ldots,W_{10}$, such that, for $t=6,12,18$ and $g=1,\ldots,10$, the response $Z_{tg}$ is distributed according to a Gamma-compound Poisson distribution with conditional expectation given by
\begin{align*}
	\log(\E[Z_{tg}\vert W_{g}=w]) &= \beta_{t}+w, \quad \text{ for all } w\in \R.
\end{align*}
Note that the Gamma-compound Poisson family of distributions is an exponential dispersion model (see \citeauthor{Jorgensen1987},\citeyear{Jorgensen1987}); therefore, the model we are defining is a genuine GLMM. Furthermore, the distributions in the Gamma-compound Poisson family have the peculiarity of attributing positive probability to the value zero and otherwise being a continuous distribution taking positive values, making them suitable for describing the lesion area.

The Gamma-compound Poisson family has been known for a long time (see \citeauthor{Tweedie1984}, \citeyear{Tweedie1984}; 
\citeauthor{Jorgensen1987},\citeyear{Jorgensen1987}, and 
\citeauthor{Cordeiro2021},\citeyear{Cordeiro2021}),
however, these distributions are not routinely used  in applications of generalised linear models (or GLMMs). Therefore, we shortly describe the inference procedure we used (for a detailed study see \citeauthor{Labouriau2021} ,\citeyear{Labouriau2021}). 
The Gamma-compound Poisson family is characterised by having a power variance function (\ie, a function expressing the variance as a function of the expectation) of the form $V(\mu) = k \mu^p$ for $p$ in the open interval $(1,2)$ (see \citeauthor{Cordeiro2021},\citeyear{Cordeiro2021}), where different power indices $p$ yield different Gamma-compound Poisson families. 
The probability of observing a zero value can be calculated as a function of the power index $p$ and the expectation (see \citeauthor{Jorgensen1987},\citeyear{Jorgensen1987}). In the analysis described above, we calculated the probability of each observation taking the value zero using a grid of values of the power index $p$. We estimated the expected number of zeroes for each observation week by summing the probability of observing a zero for each observation made in this week. This process was repeated for each value of the power index $p$ in a grid of possible values. We used in the analysis the value of the power index $p$ that minimised the Euclidean distance between the vector containing the observed proportions of zeroes for each week and the vector of expected number of zeroes for each week.

\section{Multivariate Simultaneous Models for  Responses of Different Statistical Nature}\label{Section.3}

\subsection{A Multivariate Construction}
\label{SubSection.3.1}

We use now the  marginal GLMMs described above to formulate a MGLMM that simultaneously describe the $16$ responses observed in this experiment. The idea explored here is to combine the marginal models into a MGLMM by constructing a $16$-dimensional multivariate Gaussian random component (corresponding to the $16$ responses) for each glass. More precisely, define, for $g=1,\ldots,10$,
\begin{align*}
 \bm{B}_g = \left ( U_g^{[1]},\ldots,U_g^{[14]},V_g,W_g \right ). 
\end{align*}
According to the MGLMM we define here, the random components $\bm{B}_1 , \ldots , \bm{B}_{10}$ are independent and multivariate normally distributed with distribution given by
\begin{align*}
\bm{B}_g \sim \mbox{N}_{16} \left ( \bm{0}, \bm{\Sigma} \right ),
\mbox{ for } g=1,\ldots,10 
\, .
\end{align*}
Moreover,  we assume that the multivariate vectors of  responses,
$( X_{tg}^{[1]},  \ldots   X_{tg}^{[14]}, Y_{tg} , Z_{tg} )$,  for $t=6,12,18$ and $g=1,\ldots,10$,
 are conditionally independent given the random components  $\bm{B}_1 , \ldots , \bm{B}_{10}$.
 Moreover, according to the MGLMM, for $t=6,12,18$ and $g=1,\ldots,10$, the vector of  responses
$( X_{tg}^{[1]},  \ldots   X_{tg}^{[14]}, Y_{tg} , Z_{tg}  )$ is conditionally distributed given $\bm{B}_g$ as specified below
\begin{align}\label{Eqn.3.1.1}
	\left \{
	\begin{array}{l}
		X_{tg}^{[1]} \vert U_{g}^{[1]} = u_1 
		\sim \text{Ga} \left ( \exp \left \{ \theta_{t}^{[1]} + u_1  \right \}, \lambda_1  \right ) 
		,\, \forall u_1\in\re 
		\\
		\vdots
		\\
		X_{tg}^{[14]} \vert U_{g}^{[14]} = u_{14} 
		\sim \text{Ga} \left ( \exp \left \{ \theta_{t}^{[14]} + u_{14}  \right \}, \lambda_{14}  \right ) 
		,\, \forall u_{14}\in\re 
		\\
		Y_{tg} \vert V_{g} = v 
		\sim \text{Bi} \left (9, \logitInv \left \{ \alpha_{t} + v  \right \} \right ) 
		,\, \forall v \in\re 
		\\
		Z_{tg} \vert W_{g} = w 
		\sim \text{ComPo} \left ( \exp \left \{ \beta_{t} + w  \right \}, \lambda_{Z}  \right ) 
		,\, \forall w\in\re. 
	\end{array}
	\right.
\end{align}
Here the notation $X\sim$ Ga $(\mu, \lambda)$ and $Z\sim$ ComPo $(\mu, \lambda)$ indicates that $X$ and $Z$ are distributed according to the Gamma and Gamma-compound Poisson distributions with mean $\mu$ and dispersion parameter $\lambda$, respectively. $Y\sim$ Bi $(n,p)$ denotes that $Y$ is binomially distributed with size $n$ and probability parameter $p$.

\subsection{The Covariance Structure of the Random Components}
\label{SubSection.3.2}

The covariance structure of the random components (given by the covariance matrix $\bm{\Sigma}$) will be characterised using the tools of graphical models. Before embracing this task, we give a short account of the basic theory of graphical models required for the exposition
(see \citeauthor{Whittaker1990}, \citeyear{Whittaker1990} and \citeauthor{Lauritzen1996}, \citeyear{Lauritzen1996} for details). 

Let $\UG$ denote an undirected graph with vertices composed of random 
variables. A pair of vertices belong to the set of edges $\mathcal{E}\subseteq \mathcal{V}\times \mathcal{V}$ if, and only if, the corresponding variables are conditionally dependent given the remaining variables in the set of vertices $\mathcal{V}$. Usually, we represent the graph $\UG$ by a set of points in the plane corresponding to the vertices in $\mathcal{V}$; an edge connecting two vertices is represented by a line connecting the two points corresponding to the vertices. The following basic definitions of graph theory will be necessary to characterise the covariance structure of the MGLMM we work with.
We say that there is a path connecting two vertices, say $v_1$ and $v_n$, if there exists a sequence of vertices $v_1, \ldots, v_n$ such that, for $i = 1, \ldots , n-1$, the pair $(v_i, v_{i+1})$ is in $\mathcal{E}$. A set of vertices $S$, separates two disjoint sets of 
vertices $A$ and $B$ in the graph $\UG$ when every path connecting a vertex in $A$ to a vertex in $B$ necessarily contains a vertex in $S$.  According to the theory of graphical models (see 
\citeauthor{Lauritzen1996},  \citeyear{Lauritzen1996} and \citeauthor{Perl2009}, \citeyear{Perl2009}), the graph defined above satisfies the \emph{separation principle}, which states that if a set of vertices $S $, separates two disjoint subsets of 
vertices $A $ and $B$ in the graph $\UG$, then all variables in 
$A$ are independent of all variables in $B$ given $S$.
Moreover, if the subsets $A$ and $B$ 
are isolated (\ie, there are no paths connecting a vertex in $A$ to a vertex in $B$), then the variables in $A$ are independent of  the variables in $B$.

We characterise the covariance structure of the MGLMM defined in Section \ref{SubSection.3.1} by defining a graphical model constructed with its random components. In this way, for each $g$ in $\{1,\ldots,10\}$, we might construct the graph $\mathcal{G}_g=(\mathcal{V}_g,\mathcal{E}_g)$ with the set of vertices
$\mathcal{V}_g = \{ U_g^{[1]}, \ldots  ,U_g^{[14]},V_g,W_g  \}$
using the conventions defined above.
Since the random vectors, $(U_g^{[1]},\ldots,U_g^{[14]},V_g,W_g )$ for $g=1,\ldots,10$, are per construction independent and identically distributed, the graphs $\mathcal{G}_1, \ldots \mathcal{G}_{10} $ are identical; therefore, we suppress the subindex $g$ in the discussion below and use the notation $\UG$ to refer to a generic graph representing the (common) covariance structure of the random components. 

Suppose that the multivariate random components of the MGLMM have a covariance structure encoded by the graph $\UG$. This covariance structure allows us to draw conclusions on the dependence of the unobservable random components, which is not our primary interest. In order to extend those conclusions to the observed responses, we should use the \emph{induced separation principle}, defined in \cite{Pelck2020B}, which states that if two disjoint sets of random components, for example $A=\{U^{[5]},\ldots,U^{[14]}\}$ and $B=\{V,W\}$, are conditional independent given a separating set of random components $S=\{U^{[1]},\ldots,U^{[4]}\}$, then the corresponding set of conditional responses $\tilde{A}=\{X^{[5]},\ldots,X^{[14]}\}$ and $\tilde{B}=\{Y,Z\}$ are conditional independent given the set of random components $S$. This result implies that the knowledge of the random components in $S$ renders the VOC's in $\tilde{A}$ uninformative with respect to the lesion area and the proportion of apples presenting visible symptoms. 

In the analysis of the experiment described above, we adjusted the GLMMs introduced in Section \ref{Section.2} using the Laplace approximation method proposed by Breslow and Clayton (1993).
We modelled the predicted values of the random components of those models by finding the graph which minimises the BIC (Bayesian Information Criterium) as exposed in \cite{Abreu2010} (see also \citeauthor{Edwards2010}, \citeyear{Edwards2010}).

\section{Results}
\label{Section.4}

We give below a brief description of the results obtained, see \cite{Holthusen2021a} for a full discussion.
Figure \ref{TheFigure} displays the representation of the estimated graph describing the covariance structure of the random components of the MGLMM we adjusted. 
It is remarkable that a group of only four random components related to VOCs (composed by anisole, 3-pentanone, 2-methyl-1-propanol and 2-phenylethanol) separates the random components associated to the two infection responses (\ie, the infection proportion and the lesion area) from the random components connected with the other VOCs. Therefore, by the extended separation principle, the knowledge of the random components corresponding to the concentration of  anisole, 3-pentanone, 2-methyl-1-propanol and 2-phenylethanol renders the concentrations of the other VOCs independent of the infection proportion and the lesion area.

We verified the adequacy of the MGLMM described in the following way. First, we checked the marginal GLMMs by plotting the Pearson residuals against the fitted values (not shown). No anomalies were encountered.
Moreover, we applied the cumulative distribution function of the putative distribution to each observation (with the estimated mean and dispersion) and verified whether the resulted transformed observations adhered to the uniform distribution in the interval between $0$ and $1$. All the p-values found were larger than $0.10$.

\thispagestyle{empty}
\definecolor{ashgrey}{rgb}{0.6, 0.65, 0.61}
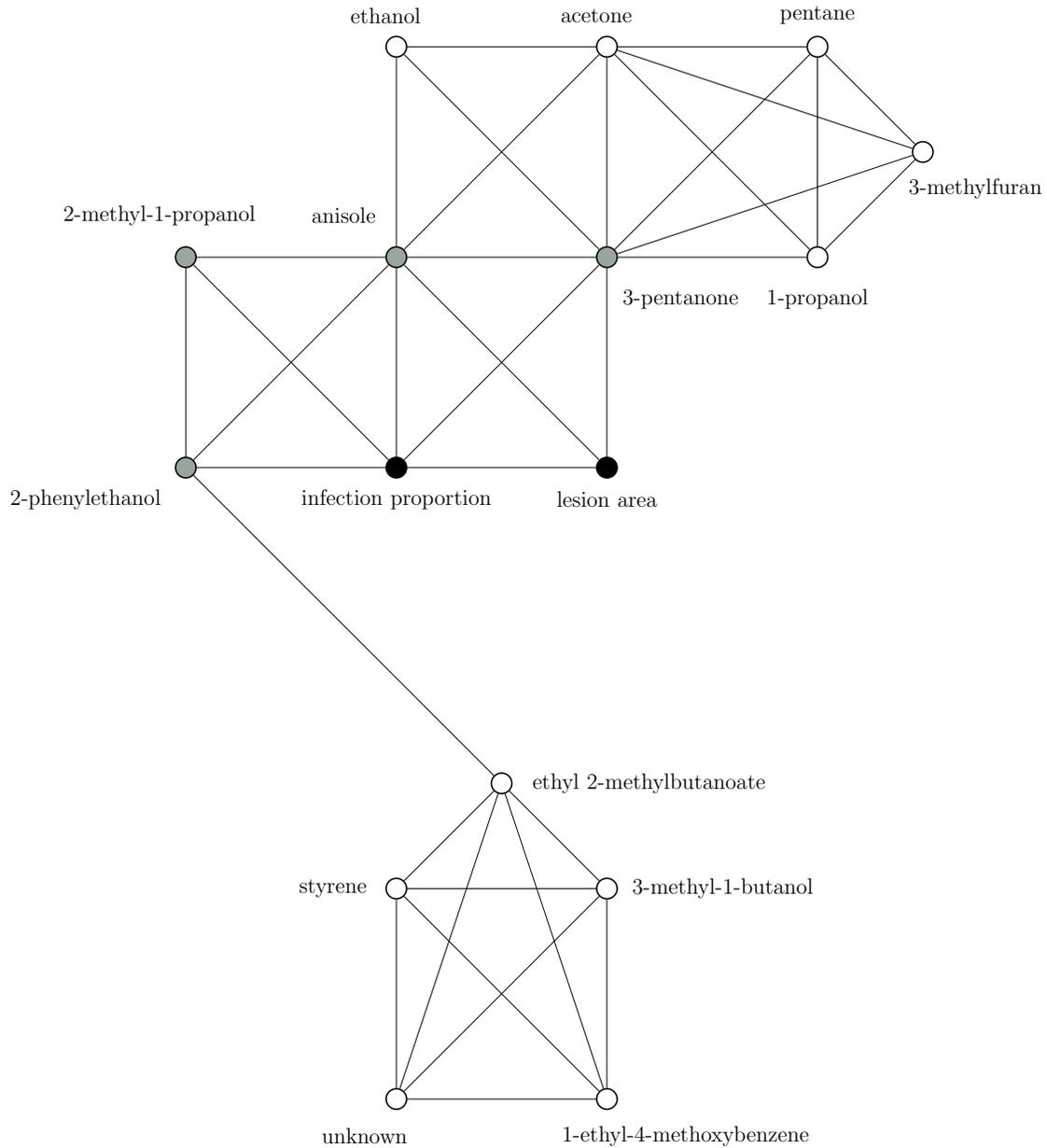
\begin{figure}[htbp]
    \begin{center} \scalebox{0.75}{
                                  \begin{tikzpicture}[
       roundnodeIII/.style={circle, draw=black, fill=ashgrey, thick},
       roundnodeII/.style={circle, draw=black, fill=black, thick},
       roundnode/.style={circle, draw=black, thick},
]
\node [style=roundnodeII] (V1) at (4, 4)   {   }; \node  at (4, 3.4) {infection proportion};

\node [style=roundnodeII] (U1) at (8, 4)   {   };  \node  at (8, 3.4) {lesion area};

\node [style=roundnodeIII] (V2) at (4, 8)   {   };  \node  at (3, 8.8) {anisole};
\node [style=roundnodeIII] (U2) at (8, 8)   {   };  \node  at (9.4, 7.2) {3-pentanone};

\node [style=roundnode] (V3) at (4, 12) {   };  \node  at (3.8,12.6) {ethanol}; 
\node [style=roundnode] (U3) at (8, 12) {   };  \node  at (7.8,12.6) {acetone}; 

\node [style=roundnodeIII] (Z2) at (0, 4) {  };  \node  at (-1.9, 3.4) {2-phenylethanol};
\node [style=roundnodeIII] (Z3) at (0, 8) {  };  \node  at (-0.5, 8.8) {2-methyl-1-propanol};

\node [style=roundnode] (A1) at (12, 8) {  };  \node  at (12,7.2) {1-propanol}; 
\node [style=roundnode] (A2) at (12, 12) {  };   \node  at (12,12.6) {pentane}; 

\node [style=roundnode] (B2) at (14, 10) {  };   \node  at (15,9.3) {3-methylfuran};

\node [style=roundnode] (Z1) at (6, -2) {  };  \node  at (8.8, -2) {ethyl 2-methylbutanoate};
\node [style=roundnode] (Z0) at (4, -4) {  }; \node  at (2.8, -4) {styrene};
\node [style=roundnode] (W0) at (4, -8) {  }; \node  at (3.4, -8.7) {unknown};
\node [style=roundnode] (W1) at (8, -8) {  };  \node  at (9.5, -8.7) {1-ethyl-4-methoxybenzene};
\node [style=roundnode] (W2) at (8, -4) {  };  \node  at (10.2, -4) {3-methyl-1-butanol};

  \draw [-] (V1) to (U1);
  \draw [-] (V2) to (U2);
  \draw [-] (V3) to (U3);
  \draw [-] (V1) to (V2);
   \draw [-] (V2) to (V3);
   \draw [-] (U1) to (U2);
   \draw [-] (U2) to (U3);
   \draw [-] (V1) to (U2);
   \draw [-] (U1) to (V2);
   \draw [-] (V2) to (U3);
   \draw [-] (V3) to (U2);
   \draw [-] (Z3) to (V2);
   \draw [-] (Z3) to (Z2);
   \draw [-] (Z2) to (V1);
   \draw [-] (Z3) to (V1);
   \draw [-] (Z2) to (V2);
   \draw [-] (U3) to (A2);
   \draw [-] (U3) to (B2);
   \draw [-] (U3) to (A1);
   \draw [-] (A2) to (B2);
   \draw [-] (A2) to (A1);
   \draw [-] (A2) to (U2);
   \draw [-] (B2) to (A1);
   \draw [-] (B2) to (U2);
   \draw [-] (A1) to (U2);
   
   \draw [-] (Z2) to (Z1);
   \draw [-] (Z1) to (Z0);
   \draw [-] (Z1) to (W0);
   \draw [-] (Z1) to (W1);
   \draw [-] (Z1) to (W2);
   \draw [-] (Z0) to (W0);
   \draw [-] (Z0) to (W1);
   \draw [-] (Z0) to (W2);
   \draw [-] (W0) to (W1);
   \draw [-] (W0) to (W2);
   \draw [-] (W1) to (W2);

  \end{tikzpicture}
}
\end{center}     
\caption{Graphical model representing the covariance structure of the random components related to the $14$ VOCs, the lesion area and the proportion of infection. The vertices related to the infection responses (lesion area and proportion of infection) are represented as black circles. The vertices directly connected to the infection responses (depicted as grey circles) separate the vertices related to the infection responses from the vertices related to the other VOCs (represented as write circles).}   
\label{TheFigure}        
\end{figure}

\clearpage
\section*{Acknowledgements}
\Fnine
The first and the last authors were partially financed by the Applied Statistics Laboratory (aStatLab) at the Department of Mathematics, Aarhus University. 
 \normalsize


\end{document}